\documentclass[aps,prl,floatfix,showpacs,twocolumn, preprintnumbers]{revtex4-2}

\usepackage{caption}
\usepackage{subcaption}
\usepackage{graphicx}
\usepackage{amssymb}
\usepackage{amsmath}
\usepackage{cancel}
\usepackage{placeins}
\usepackage[dvipsnames,usenames]{xcolor}
\usepackage{dcolumn}
\usepackage{bm}
\usepackage{physics}
\usepackage{hyperref}
\usepackage{float}
\usepackage{soul}

\usepackage[utf8]{inputenc}

\usepackage{bm}
\usepackage{color}
\usepackage{graphicx,epsfig}
\usepackage{amsfonts,amssymb,amsmath}
\usepackage{float}
\usepackage{xcolor}
\usepackage{color}
\usepackage{multirow}
\usepackage{subcaption}

\usepackage{dcolumn}     

\newcommand{\nc}{\newcommand}

\newcommand{\beq}{\begin{equation}}
\newcommand{\eeq}{\end{equation}}
\nc{\bfx}{{\bf x}}
\nc{\bfy}{{\bf y}}
\nc{\bfz}{{\bf z}}
\nc{\bfxh}{{\bf \hat{x}}}
\nc{\bfyh}{{\bf \hat{y}}}
\nc{\bfzh}{{\bf \hat{z}}}
\nc{\bfj}{{\bf j}}
\nc{\bfr}{{\bf r}}
\nc{\bfR}{{\bf R}}
\nc{\bfk}{{\bf k}}
\nc{\bfq}{{\bf q}}
\nc{\bfp}{{\bf p}}
\nc{\bfv}{{\bf v}}
\nc{\bfs}{{\bf s}}
\nc{\bfA}{{\bf A}}
\nc{\bfJ}{{\bf J}}
\nc{\bfsg}{{\bm \sigma}}
\nc{\bfta}{{\bm \tau}}
\nc{\bfvh}{{\bf \hat{v}}}
\nc{\bfqh}{{\bf \hat{q}}}
\nc{\low}{\delta_{\rm Low}}

\nc{\swap}{\rightleftharpoons}

\def\beq{\begin{equation}}
\def\eeq{\end{equation}}
\def\beqy{\begin{eqnarray}}
\def\eeqy{\end{eqnarray}}

\begin{document}

\title{Quantum Monte Carlo calculations of magnetic form factors in light nuclei}
\author{G. \ Chambers-Wall$^{1}$}
\email{chambers-wall@wustl.edu}
\author{A. \ Gnech$^{2,3}$}
\email{agnech@odu.edu}
\author{G. B. \ King$^{1}$}
\email{kingg@wustl.edu}
\author{S.\ Pastore$^{1,4}$}
\email{saori@wustl.edu}
\author{M.\ Piarulli$^{1,4}$}
\email{m.piarulli@wustl.edu}
\author{R.\ Schiavilla$^{2,3}$}
\email{schiavil@jlab.org}
\author{\mbox{R. B.\ Wiringa$^5$}}
\email{wiringa@anl.gov}

\affiliation{
$^1$\mbox{Department of Physics, Washington University in Saint Louis, Saint Louis, MO 63130, USA}\\
$^2$\mbox{Department  of  Physics,  Old  Dominion  University,  Norfolk,  VA  23529}\\
$^3$\mbox{Theory  Center,  Jefferson  Lab,  Newport  News,  VA  23610}\\
$^4$\mbox{McDonnell Center for the Space Sciences at Washington University in St. Louis, MO 63130, USA}\\
$^5$\mbox{Physics Division, Argonne National Laboratory, Argonne, IL 60439}\\
}

\begin{abstract}
We present Quantum Monte Carlo calculations of magnetic form factors in $A=6-10$ nuclei, based on Norfolk two- and three-nucleon interactions, and associated one- and two-body electromagnetic currents. Agreement with the available experimental data for $^6$Li, $^7$Li, $^9$Be and $^{10}$B up to values of momentum transfer $q\sim 3$ fm$^{-1}$ is achieved when two-nucleon currents are accounted for. We present a set of predictions for the magnetic form factors of $^7$Be, $^8$Li, $^9$Li, and $^9$C. In these systems, two-body currents account for $\sim40-60\%$ of the total magnetic strength. Measurements in any of these radioactive systems would provide valuable insights on the nuclear magnetic structure emerging from the underlying many-nucleon dynamics. A particularly interesting case is that of $^7$Be, as it would enable investigations of the magnetic structure of mirror nuclei. 
\end{abstract}
\maketitle

Electron-nucleus scattering is a powerful investigative tool to probe nuclei~\cite{Forest1966,Donnelly1984}. This is primarily due to the fact that electrons hardly perturb the nucleus, leaving us with relatively simple expressions for the cross sections, that are factorized into a nuclear structure contribution and one for the well-known physics of the external probe. Additionally, electromagnetic data are frequently determined with great accuracy, providing stringent constraints on the nuclear models. Thus, electron-nucleus scattering is a testing ground for microscopic theories modeling nuclei from the underlying many-nucleon dynamics.

The microscopic theoretical approach, describing the nucleus as a collection of nucleons interacting among each other via effective two- and three-nucleon forces, has reached remarkable sophistication~\cite{Hergert:2020bxy}. Consistent comparisons with available experimental data have repeatedly confirmed the validity of this theoretical approach. Concurrent with the development of advanced computational many-body methods, the formulation of many-nucleon interactions and currents have progressed tremendously over the last three decades. Most notably, present state-of-the-art microscopic nuclear calculations adopt chiral effective field theories ($\chi$EFTs)~\cite{Weinberg:1978kz} to construct many-nucleon interactions~\cite{Weinberg:1990rz,Weinberg:1991um,Epelbaum:2008ga,Machleidt:2017vls,Piarulli:2022hml} and currents~\cite{Park:1995pn,Kolling:2009iq,Kolling:2011mt,Krebs:2016rqz,Krebs:2019aka,Pastore:2008ui,Pastore:2009is,Pastore:2011ip,Piarulli:2012bn,Schiavilla:2018udt,Gnech:2022vwr} that are grounded in the underlying theory of quantum chromodynamics (QCD)~\cite{Gross:2022hyw}. Chiral effective field theory is a low energy (and low momentum) approximation of QCD that preserves the symmetries exhibited by the fundamental theory below a breaking scale $\Lambda_\chi\sim1$ GeV. Within this formalism, many-nucleon operators are arranged in a power expansion of the low-momentum scale over $\Lambda_{\chi}$. This allows for a determination of the theoretical uncertainty associated with neglecting sub-leading terms in the chiral expansion. The $\chi$EFT framework has been adopted primarily to study low-energy observables, including binding energies and spectra~\cite{Piarulli:2017dwd,Ekstrom:2015rta,Hergert:2015awm,Lynn:2017fxg,Simonis:2017dny,Lonardoni:2018nob,Jiang:2020the}, electromagnetic moments~\cite{Gnech:2022vwr,Martin:2023dhl,Pal:2023gll,Miyagi:2023zvv,Lechner:2023lyr}, and transitions~\cite{Pastore:2012rp,Pastore:2014oda,Parzuchowski:2017wcq,Stroberg:2022ltv,McCoy:2024kah}. Analysis of the validity of $\chi$EFT at higher values of momentum transfer has been conducted principally via the study of electromagnetic form factors of the deuteron and the trinucleon systems, $^3$H and $^3$He~\cite{Phillips:2006im,Valderrama:2007ja,Song:2008zf,Kolling:2012cs,Piarulli:2012bn,Gnech:2022vwr}. The predictive power of this theory beyond the low-energy/low-momentum regime in heavier nuclei remains to be understood. Because this regime is critical to interpret experiments measuring superallowed $\beta$-decay~\cite{Seng:2022cnq,Cirigliano:2024rfk,Gennari:2024sbn}, neutrinoless double beta decay~\cite{Engel:2016xgb,Agostini:2022zub,Pastore:2017ofx,Cirigliano:2019vdj}, and long-baseline neutrino oscillation experiments~\cite{Abe:2019,NOvA:2019cyt,Acciarri:2017,Aliaga:2014,Seo:2018,Abi:2020,Ruso:2022qes}, it is important verify that the model is valid for these kinematics. 

 In this letter,  we address the aforementioned question  by focusing on the magnetic structure of $A=6-10$ nuclei, with emphasis on magnetic form factors. Specifically, we perform our calculations using two different implementations of Quantum Monte Carlo computational methods~\cite{Carlson2015}, namely variational Monte Carlo (VMC) and Green's function Monte Carlo (GFMC). These are two stochastic approaches to solve the Schr\"odinger equation of strongly correlated nucleons that incorporate the complexity emerging from two- and three-nucleon interactions.

We base our results on the Norfolk two- and three nucleon interactions~\cite{Piarulli:2014bda,Piarulli:2016vel,Piarulli:2017dwd,Baroni:2018fdn,Piarulli:2019cqu} that are derived up to N3LO in the chiral expansion from a $\chi$EFT that retains pions, nucleons, and intermediate $\Delta$'s as explicit degrees of freedom. The long- and intermediate-range parts of the nucleon-nucleon interaction are modeled with one-pion and two-pion exchange mechanisms, respectively, while short-range dynamics are encoded in contact-like interactions with strengths specified by Low Energy Constants (LECs). The latter are obtained from fits to nucleon-nucleon scattering data~\cite{Piarulli:2014bda,Piarulli:2016vel,Piarulli:2017dwd,Baroni:2018fdn,Piarulli:2019cqu,Perez:2013jpa,Perez:2013oba,Perez:2014yla}, characterized by a $\chi^2$ per datum close to one. Similarly, three-nucleon interactions involve pion-exchange-like mechanisms and contact terms, involving this time two LECs. In the present study, these are fit to reproduce the trinucleon biding energies, along with the triton beta decay half-life~\cite{Piarulli:2014bda,Piarulli:2016vel,Piarulli:2017dwd,Baroni:2018fdn,Piarulli:2019cqu}. Norfolk models based on this fitting strategy are denoted in the literature with `NV2+3$^\star$'. Here, we will discuss results based on two implementations of the `NV2+3$^\star$' model, namely `NV2+3-Ia$^\star$' and `NV2+3-IIb$^\star$'. These models differ in the nucleon-nucleon scattering data sets adopted to constrain the two-nucleon interaction, and in the choice of regulators inherent to the theory~\cite{Piarulli:2014bda,Piarulli:2016vel,Piarulli:2017dwd,Baroni:2018fdn,Piarulli:2019cqu}. 

The interaction of the external electromagnetic probe with individual nucleons and with pairs of correlated nucleons is described by one- and two-nucleon currents, respectively. The one-body current, appearing at leading order in the chiral expansion, consists of the standard convection and the spin-magnetization currents associated with an individual nucleon. In the present study, we account for additional (and negligible) relativistic corrections to this operator, which enter at N2LO in the chiral expansion. Electromagnetic two-body currents have been extensively studied within several implementations of $\chi$EFT~\cite{Park:1995pn,Kolling:2009iq,Kolling:2011mt,Krebs:2016rqz,Krebs:2019aka,Pastore:2008ui,Pastore:2009is,Pastore:2011ip,Piarulli:2012bn,Schiavilla:2018udt,Gnech:2022vwr}. Here, we adopt electromagnetic currents that are consistent with the Norfolk interactions that encompass the same nucleon-nucleon dynamics as those present in the Norfolk two-nucleon potential. These currents have been most recently summarized in Refs.~\cite{Schiavilla:2018udt,Gnech:2022vwr}, where they are evaluated up to N3LO in the chiral expansion. Two-body electromagnetic currents consist of one- and two-pion range contributions, as well as contact terms encoding short range dynamics. They involve five unknown LECs which are determined to fits to the deuteron and trinucleon magnetic moments, as well as deuteron threshold electrodisintegration at backward angles~\cite{Gnech:2022vwr}. We note that the dominant two-nucleon current appears at NLO in the chiral expansion and consists of the well-known seagull and pion-in-flight contributions of one-pion range~\cite{RISKA1970662,Chemtob:1971pu,Riska:1989bh,Schiavilla:1990yj,Marcucci:2005zc}. The NLO current is purely isovector in nature. 

\begin{figure}[tbh]
\begin{center}
    \includegraphics[width=0.45\textwidth]{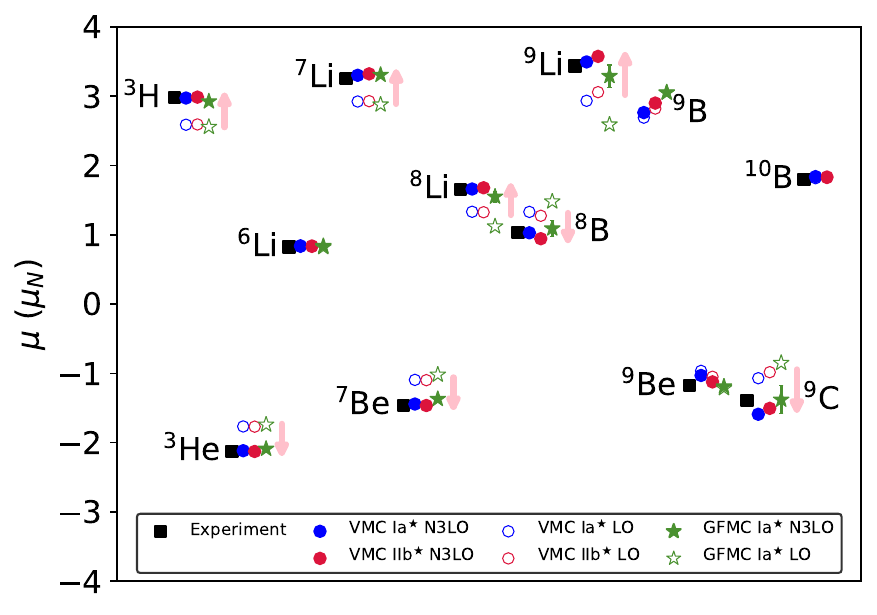}
\end{center}
\caption{VMC magnetic moments calculated with the NV2+3-Ia$^{\star}$ (blue circles) and NV2+3-IIb$^{\star}$ (red circles) compared with experiment (black dots). GFMC results based on NV2+3-Ia$^{\star}$ are given by the green stars. Empty symbols denote the calculation up to LO in the current, while filled symbols include all contributions through N3LO.}\label{fig:exp.comp}
\end{figure}

{\it Magnetic moments} -- 
We test the NV2+3$^\star$ interaction and current models at zero value of momentum transfer by evaluating the magnetic moments of light nuclei. These calculations are summarized in Fig.~\ref{fig:exp.comp}, where the experimental data are shown by black squares, while VMC and GFMC results are shown by circles and stars, respectively. Calculations based on the one-body operator alone are indicated with empty symbols, those that include two-body currents up to N3LO are indicated with filled symbols. Pink arrows indicate the major shifts induced by the two-body currents, which are positive (negative) for neutron-rich (proton-rich) nuclei. Since the deuteron and trinucleon magnetic moments have been used to constrain the two-body current operator, predictions are solely for the $A=6-10$ nuclei. Two-body currents give a negligible correction to the LO magnetic moment in $^6$Li and $^{10}$B. These systems have isospin $T=0$, therefore the two-body current of one-pion range at NLO does not contribute to their magnetic moments. While the calculations at LO offer a qualitative explanation of the data, the inclusion of two-body currents brings the theory in agreement with the experiment. In $^9$B and $^9$Be, the suppression of two-body current contributions is understood by analyzing their specific structures. These systems exhibit (predominantly) a $2\alpha+N$ cluster structure, where $N$ is the valence nucleon, which is driving the total magnetic moment of the system. Conversely, $^9$C and $^9$Li are predominantly in the $\alpha + ^3{\rm He} + pp$ and $\alpha + ^3{\rm H} + nn$ cluster structures, respectively. Here, two-body currents are active within the trinucleon clusters, and between the trinucleon clusters and the valence nucleons, leading to large contributions in these systems. These findings are consistent with previous studies performed within the QMC approach~\cite{Pastore:2012rp,Martin:2023dhl}

{\it Magnetic form factors} -- The microscopic framework has been extensively utilized to study magnetic form factors of $A=2$ and $3$ nuclei, both within phenomenological~\cite{Carlson:1997qn} and $\chi$EFT-based approaches~\cite{Phillips:2006im,Valderrama:2007ja,Song:2008zf,Kolling:2012cs,Piarulli:2012bn}. To the best of our knowledge, for nuclei beyond $A=3$, the microscopic approach has been applied only to  $^6$Li. This system has been studied in Ref.~\cite{Wiringa:1998hr}, where the VMC results are based on the Argonne (AV18)~\cite{Wiringa:1994wb} and Urbana-IX (UIX)~\cite{Pudliner:1995wk}, two- and three-nucleon interactions, respectively, along with  consistent two-body meson-exchange currents. 

 Here, we report the first microscopic calculation of magnetic form factors of $A=6-10$ nuclei based on a $\chi$EFT formulation. Because of the minimal model dependence on observables when including two-body currents through N3LO (see Ref.~\cite{prc} for a detailed analysis of this dependence), we present results using the Norfolk NV2+3-IIb$^\star$. For this work we use the definition given in Ref.~\cite{Donnelly1984}, {\it i.e.}, 
\begin{equation}
\begin{aligned}
|F_M(q)|^2= \frac{1}{2 J+1} \sum_{L=1}^{\infty}\left|\left\langle J ||M_L(q)|| J\right\rangle\right|^2\,,\label{eq:fm}
\end{aligned}
\end{equation}
where the magnetic form factor is expressed in terms of the reduced matrix elements of the magnetic  multipole operators ($M_L$),  $J$ is the total angular momentum of the nucleus, and $q$ is the momentum transfer. We extract the multipole operators from the matrix elements of the electromagnetic current operator, as described in Ref.~\cite{Carlson2015}. Interested readers can find the explicit derivation of the relevant expressions in the companion paper of Ref.~\cite{prc}. Finally, we note that the computational overhead to perform more accurate GFMC calculations provides minimal gain in interpreting the present experimental data; however, should more accurate measurements become available, it could merit a future GFMC study

\begin{figure}
    \includegraphics[width=3.1in]{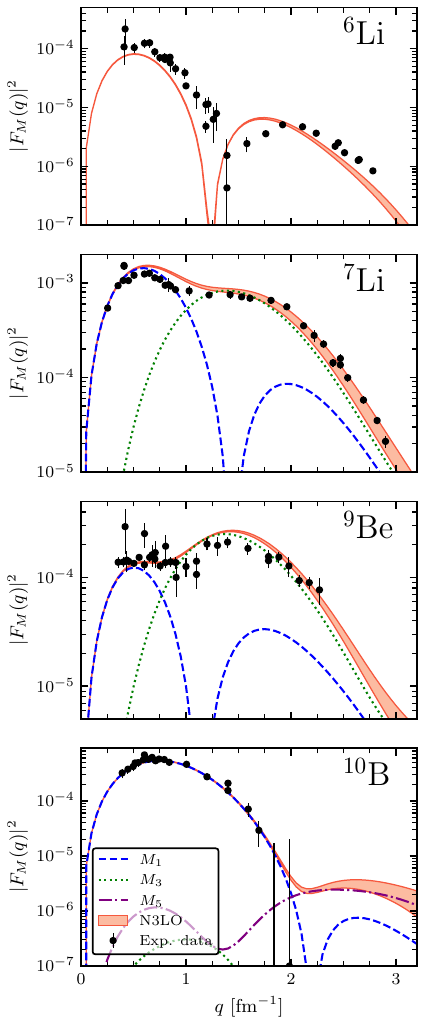}
    \caption{VMC calculations of the $^6$Li, $^7$Li, $^9$Be, and $^{10}$B magnetic form factors, computed with the NV2+3-IIb$^\star$ interaction and consistent electromagnetic current. In the figure we show the contribution of the multipoles $M_1$ (blue dashed line), $M_3$ (green dotted line) and $M_5$ (purple dot dashed line) computed at N3LO. 
    Results accounting for the full current at N3LO are denoted by the red band. The error band reflects theoretical uncertainties arising from leaving out terms beyond N3LO in the  currents. 
}\label{fig:ff.expt}
\end{figure}

In Fig.~\ref{fig:ff.expt}, we present the results for nuclei for which experimental data are available. The experimental data (black circles) are very limited and affected by large uncertainties. Indeed, the majority of them were taken before 1984, as it can be appreciated in  Table~\ref{tab:data}, where we compile the presently available experimental data. Magnetic form factors, calculated using the full current at N3LO are denoted by the red band. The error band reflects theoretical uncertainties arising from leaving out terms beyond N3LO in the  currents computed using the approach of Ref.~\cite{Epelbaum:2014efa,Epelbaum:2014sza}. Additionally, we show the contributions from the various multipoles, with $M_1$, $M_3$, and $M_5$ indicated by dashed blue, dotted green, and dashed dotted purple lines, respectively.
The calculations reproduce nicely both the position and the strength of the first peak of the magnetic form factors of $^7$Li, $^9$Be and $^{10}$B. For $^7$Li and $^9$Be, the $M_3$ contribution fills  the $M_1$'s diffraction minimum, and generates the second peak appearing in the magnetic form factors. For both nuclei, the contribution of two-body currents is crucial to reproduce the strength of the peaks, even though some extra strength is generated in the first peak of $^7$Li and in the second peak of $^9$Be. This is very similar to the findings 
of Ref.~\cite{Donnelly1984}, where less sophisticated 
nuclear structure models have been used.
In the case of $^{10}$B, the diffraction minimum is filled by the $M_5$ multipole. Note that, while this was hypothesized in Ref.~\cite{Donnelly1984}, this work presents the first confirmation of this hypothesis. 
In the case of $^6$Li, the $M_3$ multipole is not allowed for symmetry reasons. The diffraction minimum exhibited by the $M_1$ multipole is reproduced well by the calculations. However, the calculated strengths and positions of the peaks fail to reach full agreement with the data. This is possibly attributable to deficiencies in the adopted many-body nuclear model. Indeed, the description of the data improves when using the Norfolk NV2+3-IIa$^\star$ interaction, and it is almost perfect in calculations based on the AV18+UIX model, as it is found in Ref.~\cite{Wiringa:1998hr}.  In this case, the contribution of two-body currents is minimal, as expected for isoscalar nuclei.

\begin{figure}
    \includegraphics[width=3.1in]{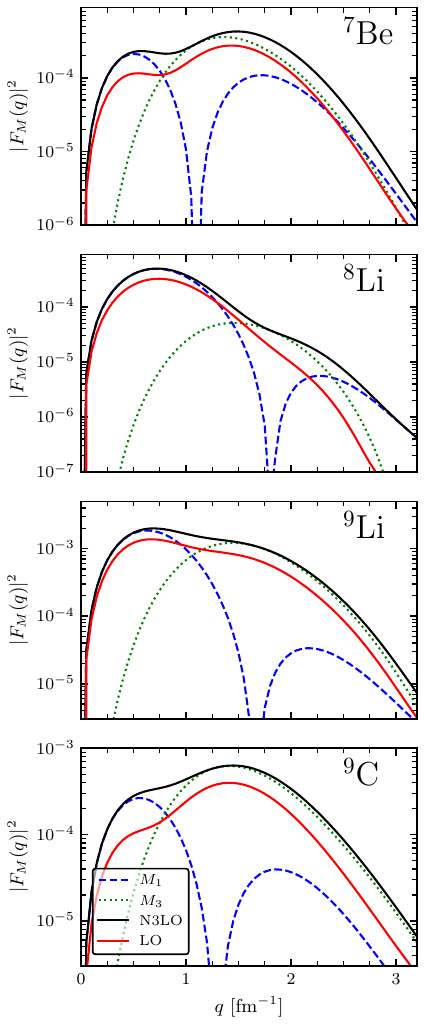}
    \caption{VMC calculations of magnetic form factors computed with the NV2+3-IIb$^\star$ interaction and consistent electromagnetic current
    for $^7$Be, $^8$Li, $^9$Li, and $^9$C. In the figure we show the contribution of the multipoles M1 (blue dashed line) and M3 (green dotted line) computed at N3LO. The full black (red) line represents the total contribution at N3LO (LO). }\label{fig:ff.no.expt}
\end{figure}

In Fig.~\ref{fig:ff.no.expt}, we show predictions for the magnetic form factors of $^7$Be, $^8$Li, $^9$Li, and $^9$C. Here, in addition to displaying the multipoles, we indicate with red and black lines calculations based on the LO one-body operator and on the full current operator at N3LO, respectively.  
For all nuclei, we consistently observe the pattern where the $M_1$ diffraction minimum is filled by the $M_3$ multipole. Note the inversion of strength of the $M_1$ and $M_3$ multipoles in the mirror nuclei, $^9$Li-$^9$C. We observe a similar inversion also in the other mirror nuclei under consideration, namely, $^7$Li-$^7$Be, and $^9$Be-$^9$B. It is worth noting that, to the best of our knowledge,  no previous experimental analysis or theoretical calculation has highlighted this phenomena. This pattern is generated by constructive (destructive) interference between the spin and orbital LO components in the $M_1$ multipole for nuclei whose magnetic moments come primarily from an unpaired proton (neutron). The orbital term is instead negligible for the $M_3$ multipole. Comparing the black and red curves in Fig.~\ref{fig:ff.no.expt}, it is evident that the contribution of the two-body currents accounts typically for $\sim40-60\%$ of the total strength in the region of $0.5\leq q\leq 3$ fm$^{-1}$. This makes all these nuclei ideal candidates for experimentally investigating two-body currents in a region of momentum transfer that is relevant to the  electron and neutrino scattering experimental programs.
In particular, we individuate $^7$Be as the best candidate for possible experimentation. Moreover, measuring the form factor of this nucleus would permit the experimental confirmation of the $M_1$-$M_3$ strength inversion in the $^7$Li-$^7$Be mirror system.

{\it Conclusions} -- In this letter, we presented QMC calculations of magnetic moments and magnetic form factors in $A=6-10$ nuclei. The study was based on the $\chi$EFT-based Norfolk two- and three-nucleon interactions, and consistent one- and two-nucleon electromagnetic currents. This framework was validated in calculations of magnetic moments that highlight the importance of two-nucleon currents to explain the data. 

The calculations of magnetic form factors were found to be in very nice agreement with the available data, correctly reproducing the strengths and the peaks characterizing the shapes of the experimental form factors. Two-body currents were again found to be relevant to reach agreement with the data in a wide range of momentum transfer.  We showed that  the missing strength in the second peak of $^{10}$B's magnetic form factor, is provided by the $M_5$ multipole, substantiating what was conjectured in Ref.~\cite{Donnelly1984}.
Comparisons with experimental data were possible only for $^6$Li, $^7$Li, $^9$Be and $^{10}$B, for which the data found in the literature dates back to the 1980's, at best. 

This study has several merits. From the theoretical point of view, this is the first set of QMC calculations of $A=7-10$ ($^6$Li was studied in Ref.~\cite{Wiringa:1998hr} using the AV18+UIX model, leading to excellent agreement with the data). Moreover, unlike the study of Ref.~\cite{Wiringa:1998hr}, the present work is based on a $\chi$EFT formulation, which facilitated the estimation of model dependencies and theoretical uncertainties arising from neglecting sub-leading terms in the currents. The remarkably good description of the data in a momentum transfer regime up to $\sim 0.6 $ GeV supports the validity of $\chi$EFT predictions well beyond the standard low-energy regime of the present applications.

Further, we provided predictions for the magnetic form factors $^7$Be, $^8$Li, $^9$Li, and $^9$C, for which there are currently no experimental data. In all of these systems, two-body currents were found to provide sizable  (up to $\sim50\%$) contributions. A pattern emerged for mirror nuclei, where the strengths of the $M_1$ and $M_3$ multipoles are found to be swapped. This feature would be very interesting to investigate further, provided that experimental data were available for a set of mirror nuclei. The predictions for these radioactive isotopes could be of relevance to the RIKEN (and other facilities') experimental program, where novel techniques have recently made it possible to study electron scattering from radioactive nuclei~\cite{Tsukada:2023}. Specifically, measurements in $^7$Be would be ideal to study both electromagnetic currents, and their effect in mirror systems. 
A detailed  discussion of the methods and results presented in this letter, can be found in an associated manuscript, submitted to PRC concurrently to this letter~\cite{prc}. In the extended paper, we provide calculations of magnetic form factors for all the nuclei appearing in Fig.~\ref{fig:exp.comp}, and for additional interactions in the NV2+3 family, along with detailed insights on the work presented here.

\begin{table}
\begin{center}
\begin{tabular}{llc}
  \hline
  \hline
Nucleus & Reference & Data type \\
\hline
  ${}^6$Li & Peterson 1962~\cite{Peterson1962} & N \\
           & Goldemberg 1963~\cite{Goldemberg1963} & N \\
           & Rand 1966~\cite{Rand1966} & N \\
           & Lapikas 1978~\cite{Lapikas1978} & D \\
           & Bergstrom 1982~\cite{Bergstrom1982} & N \\
           \hline 
    ${}^7$Li & Peterson 1962~\cite{Peterson1962} & N \\
    & Goldemberg 1963~\cite{Goldemberg1963} & N \\
    &Van Niftrik 1971~\cite{Vanniftrik1971} & D \\
    &Lichtenstadt 1983~\cite{Lichtenstadt1983} & N \\
     \hline 
    ${}^9$Be & Goldemberg 1963~\cite{Goldemberg1963} & N \\
    & Vanpraet 1965~\cite{Vanpraet1965} & N \\
    & Rand 1966~\cite{Rand1966} & N \\
    & Lapikas 1975~\cite{Lapikas1975} & N \\
     \hline 
     ${}^{10}$B & Goldemberg 1963~\cite{Goldemberg1963} & N \\
     & Goldemberg 1965~\cite{Goldemberg1965} & N \\
     &Vanpraet 1965~\cite{Vanpraet1965} & N  \\
     & Rand 1966~\cite{Rand1966} & N \\
     & Lapikas 1978~\cite{Lapikas1978} & D \\
  \hline
  \hline
\end{tabular}
\end{center}
\caption{\label{tab:data}
  Summary table of experimental data sets available for the nuclei considered in this work. In the last column we emphasize if the numerical values of the magnetic form factors (or elastic cross sections) were provided in the references (N), or if we needed to digitize from the figures (D). More details about the datasets may be found in the companion paper of Ref.~\cite{prc}.} 
\end{table}

\acknowledgments 

{\it Acknowledgments} -- This work is supported by the US Department of Energy under Contracts No. DE-SC0021027 (G.C.-W., G.B.K., and S.P.), DE-AC02-06CH11357 (R.B.W.), DE-AC05-06OR23177 (A.G. and R.S.), a 2021 Early Career Award number DE-SC0022002 (M.P.), the FRIB Theory Alliance award DE-SC0013617 (M.P.), and the NUCLEI SciDAC program (S.P., M.P., and R.B.W.). G.B.K. would like to acknowledge support from the U.S. DOE NNSA Stewardship Science Graduate Fellowship under Cooperative Agreement DE-NA0003960. G.~C.-W. acknowledges support from the NSF Graduate Research Fellowship Program under Grant No. DGE-213989. We thank the Nuclear Theory for New Physics Topical Collaboration, supported by the U.S.~Department of Energy under contract DE-SC0023663, for fostering dynamic collaborations. A.G. acknowledges the direct support of the Nuclear Theory for New Physics Topical collaboration.

The many-body calculations were performed on the parallel computers of the Laboratory Computing Resource Center, Argonne National Laboratory, the computers of the Argonne Leadership Computing Facility (ALCF) via the INCITE grant ``Ab-initio nuclear structure and nuclear reactions'', the 2019/2020 ALCC grant ``Low Energy Neutrino-Nucleus interactions'' for the project NNInteractions, the 2020/2021 ALCC grant ``Chiral Nuclear Interactions from Nuclei to Nucleonic Matter'' for the project ChiralNuc, the 2021/2022 ALCC grant ``Quantum Monte Carlo Calculations of Nuclei up to $^{16}{\rm O}$ and Neutron Matter" for the project \mbox{QMCNuc}, and by the National Energy Research
Scientific Computing Center, a DOE Office of Science User Facility
supported by the Office of Science of the U.S. Department of Energy
under Contract No. DE-AC02-05CH11231 using NERSC award
NP-ERCAP0027147.

\appendix

\bibliography{biblio,bib_magn}

\end{document}